\documentclass[preprint,aps]{revtex4}

\usepackage{graphicx}
\usepackage{dcolumn}
\usepackage{bm}
\usepackage{epsf}

\newcommand{\be}{\begin{eqnarray}}

\newcommand{\Eq}[1]{Eq.~(\ref{#1})}
\newcommand{\Eqs}[2]{Eqs.(\ref{#1},\ref{#2})}

\newcommand{\ur}[1]{(\ref{#1})}

\newcommand{\beq}{\begin{equation}}
\newcommand{\eeq}{\end{equation}}

\newcommand{\la}[1]{\label{#1}}
\newcommand{\bea}{\begin{eqnarray}}
\newcommand{\eea}{\end{eqnarray}}
\newcommand{\beqa}{\begin{eqnarray}}
\newcommand{\eeqa}{\end{eqnarray}}
\newcommand{\ba}{\begin{array}}
\newcommand{\ea}{\end{array}}

\newcommand{\half}{{\textstyle{\frac{1}{2}}}}
\newcommand{\third}{{\textstyle{\frac{1}{3}}}}

\newcommand{\n}{\nonumber}

\newcommand{\Th}{$\Theta^+\,$}

\def\appendix{\par
\setcounter{subsection}{0}
\setcounter{equation}{0}

\def\thesection{Appendix}
\def\theequation{\Alph{section}.\arabic{equation}}}

\begin{document}

\title{\bf Prediction of new charmed and bottom exotic pentaquarks}

\author{\bf Dmitri Diakonov$^{1,2}$}

\affiliation{
$^1$Petersburg Nuclear Physics Institute, Gatchina 188300, St. Petersburg, Russia \\
$^2$Yukawa Institute for Theoretical Physics, Kyoto University, Kyoto, Japan
}

\date{March 10, 2010}

\begin{abstract}

Baryons of the type $Qqqq\bar q$ (where $Q=c,b$ and $q=u,d,s$ quarks) forming anti-decapenta
($\overline{15}$)-plets with spin-parity $\half^+$ are predicted on simple theoretical considerations.
The lightest members of these multiplets are explicitly exotic doublets $cuud\bar s,\;cudd\bar s$
with mass about 2420 MeV, and $buud\bar s,\;budd\bar s$ with mass about 5750 MeV, only 130 MeV
heavier than $\Lambda_c$ and $\Lambda_b$, respectively, and thus stable against strong decays.
Although the production rate is probably very low, these remarkable pentaquarks can be looked for
at LHC, Fermilab, B-factories, RHIC and elsewhere: their signatures are briefly discussed.\\



\end{abstract}

\maketitle

\section{Introduction}

In this paper, arguments are suggested in favor of the existence of exotic pentaquarks which may well
prove to be the second lightest charmed (bottom) baryons, after $\Lambda_c$ and $\Lambda_b$. Since they
are light, the new baryons decay only weakly. They may have escaped direct observation in the past because
the production rate is expected to be quite low.

The arguments are based on considering baryons at large number of colors $N_c$. While in the real
world $N_c$ is only three, we do not expect qualitative difference in the baryon spectrum with the
large-$N_c$ limit. The bonus is that at large $N_c$ baryon physics simplifies considerably,
which enables one to take into full account the important relativistic and field-theoretic effects
that are often ignored.

The relativistic approach to baryons is key to the prediction. In implies that baryons are not
just three (or $N_c$) quarks but contain additional quark-antiquark pairs, as it is well known
experimentally. Baryon resonances may be formed not only from quark excitations as in the customary
non-relativistic quark models, but also from particle-hole excitations and ``Gamov--Teller''
transitions. At large $N_c$ these effects become transparent and tractable. At $N_c\!=\!3$ it is
a mess called ``strong interactions''. The hope is that if one develops a clear picture at large $N_c$,
its imprint will be visible at $N_c\!=\!3$.

The approach can be illustrated by the chiral quark soliton model~\cite{DP-86} or by the
chiral bag model~\cite{Hosaka} but actually the arguments of this paper are much more general.
Dynamics is not considered here, which today would require adopting a model. A concrete model would
say what is the ``intrinsic'' relativistic quark spectrum in baryons. It may get it approximately
correct, or altogether wrong. Instead of calculating the intrinsic spectrum from a model, I extract
it from the known baryon spectrum by interpreting baryon resonances as collective excitations
about the ground state and about the one-quark and particle-hole transitions.

In Section II the key question what is the symmetry of the ground-state baryon is addressed.
Arguments are presented that it is not the expected maximal possible symmetry. In particular,
$SU(3)$ flavor symmetry is spontaneously broken even in the limit of zero current quark masses.
The two critical consequences are: ({\it i}) the intrinsic spectrum of $s$ quarks in a baryon
is totally different from that of $u,d$ quarks, and ({\it ii}) the observable baryon spectrum
has characteristic ``rotational bands'' following from quantizing the rotations of the baryon
as a whole in flavor and ordinary spaces.

In Sections III, IV the lowest light baryon resonances both with positive and negative parities are
interpreted from the above viewpoint. As a byproduct, the light exotic pentaquark \Th is theoretically
confirmed at about 1520 MeV and shown to be a consequence of the existence of three well-known resonances:
$N(1440,1/2^+)$, $\Lambda(1405,1/2^-)$ and $N(1535,1/2^-)$.

In Sections V, VI heavy baryons are discussed using the intrinsic quark spectrum established from light
baryons -- at large $N_c$ the intrinsic spectra are the same in light and heavy baryons, up to $1/N_c$
corrections. Numerical checks show that the appropriate relations between light and heavy baryons
are reasonably satisfied even at $N_c\!=\!3$.

Section VII is central as an {\it anti-decapenta} ($\overline{15}$)-plet of heavy exotic charmed (bottom)
pentaquarks is predicted there. It is explained why the lightest members of that multiplet are relatively
light and hence stable under strong decays. These pentaquarks are distinct from {\em anti-}charmed (bottom)
pentaquarks suggested previously, that are about 500 MeV heavier.

In Sections VIII, IX certain properties of the predicted pentaquarks are discussed, in particular
possibilities to observe them experimentally.

The Appendix deals with the mathematical description of the ``rotational bands'' about various
intrinsic quark excitations.

\section{Mean field in baryons}

Recently a classification of baryon resonances was suggested, according to what they would look like
if the number of colors $N_c$ was large~\cite{D-09,D-09-1}. Long experience tells us that the large-$N_c$
world does not differ much from the real world with $N_c=3$, except for several very special cases,
and in many circumstances the $1/N_c$ corrections are under control~\cite{Manohar}.

At large $N_c$, the $N_c$ quarks constituting a baryon can be considered in a mean (non-fluctuating)
mesonic field which does not change as $N_c\to\infty$. Consequently, all quark levels in the mean field are
stable in $N_c$. All negative-energy levels should be filled in by $N_c$ quarks in the antisymmetric
state in color, corresponding to the zero baryon number state. Filling in the lowest positive-energy
level makes a baryon. Exciting higher quark levels or making particle-hole excitations produces
baryon resonances. The baryon mass is ${\cal O}(N_c)$, and the excitation energy is ${\cal O}(1)$.
When one excites one quark the change of the mean field is ${\cal O}(1/N_c)$ that can be neglected
to the first approximation.

The key issue is what is the symmetry of the mean field. In the chiral limit when the strange quark mass $m_s$
is set to zero, $u,d,s$ quarks are on the same footing, and one may think that the mean field is maximally
symmetric, that is flavor symmetric and spherically symmetric or, in the mathematical language, invariant under
$SU(3)_{\rm flav}\times SO(3)_{\rm space}$ rotations. Although natural, this assumption is wrong. For unknown
dynamical reasons the maximal possible symmetry of the mean field in baryons is broken spontaneously down
to the $SU(2)_{{\rm iso}\!+\!{\rm space}}$ symmetry: the mean field is invariant only under simultaneous
isospin and compensating space rotations.

The case is analogous to heavy nuclei: For some reasons many of the large-$A$ nuclei, although not all,
are not spherically-symmetric (as would seem natural) but have an ellipsoid form. It means that spherical
symmetry, or invariance under $SO(3)_{\rm space}$ rotations, is partially spontaneously broken in
the ground state. The mean field in such nuclei is not invariant under arbitrary rotation but only
with respect to rotation about the symmetry axis of the ellipsoid. In principle, symmetry could be broken
completely, {\it e.g.} a heavy nucleus could be a three-axes ellipsoid, or the mean field in baryons
could have no symmetries at all. However in practice this does not happen: the symmetry is broken but
not completely.

The question what is the symmetry of the mean field can be answered theoretically if full dynamics is
well understood: one has to try all possible symmetry patterns and check which of them leads to the lowest
energy of the ground state. It is a quantitative question. In the absence of a reliable dynamical theory
one can, however, use phenomenological, circumstantial evidence in favor of this or that symmetry.
For example, if symmetry is spontaneously broken one expects low-lying excitations, the (pseudo) Goldstone
modes, their number being equal to the number of broken symmetry generators.

If the broken symmetry group is compact (like $SU(3)_{\rm flav}$ and $SO(3)_{\rm space}$) the energy of the
Goldstone excitations is quantized. One expects then rotation bands about the ground state and about each one-particle
and particle-hole ${\cal O}(1)$ excitation, split as $1/I$ where $I$ is the moment of inertia. For heavy nuclei,
$I$ scales as $I\sim mr^2\sim A^{\frac{5}{3}}$ whereas for the baryon it scales as $I\sim N_c$ (since the baryon
radius does not rise with $N_c$). In most of heavy nuclei ($A\gg 1$) one clearly sees rotational excitations
whose splitting is much less than the ${\cal O}(1)$ one-particle and particle-hole excitations~\cite{BM}.
This is a clear evidence that such nuclei are not spherically-symmetric, otherwise there would have been
no rotational bands at all.

In real-world baryons there is no spectacular separation of scales since $N_c=3$ is not a very large number.
However some $SU(3)$ multiplets have definitely smaller splitting between themselves than others. This is an
indication that certain baryon multiplets can be interpreted as rotational states whereas others are one-particle
or particle-hole excitations. It implies, then, that the would-be $SU(3)_{\rm flav}\times SO(3)_{\rm space}$
symmetry of the mean field is broken; the question is what is the pattern.

If the mean field is only $SU(2)_{{\rm iso}+{\rm space}}$ invariant, the quantization of the rotations
needed to restore the original $SU(3)_{\rm flav}\times SO(3)_{\rm space}$ symmetry for a ground-state baryon
leads precisely to the baryon multiplets $({\bf 8},\half^+)$ and $({\bf 10},\frac{3}{2}^+)$ observed in Nature.
It is an argument in favor of this particular pattern of symmetry breaking.
To specify the $N_c$ behavior of the splitting between the centers of these multiplets, one needs to
generalize them to certain prototype $SU(3)$ multiplets at arbitrary $N_c$, that reduce to the octet and decuplet
at $N_c=3$~\cite{DuPrasz,DP-04}: the splitting turns out to be $3/2I_1={\cal O}(1/N_c)$, see the Appendix.
Numerically this splitting is 1382-1152=230 MeV, such that $1/I_1=153\,{\rm MeV}$. The number is indeed
considerably less than the splitting from the center of the next nearest $({\bf 8},\half^+)$ multiplet involving
the Roper resonance, 1630-1152=478 MeV. In the present interpretation, the first splitting is ${\cal O}(1/N_c)$
as due to the rotation of a baryon as a whole, whereas the second is ${\cal O}(1)$ and is due to a one-quark
excitation in the mean field~\cite{D-09}.

I note in passing that in the non-relativistic quark model the splitting between the lowest
octet and decuplet is interpreted as due to hyperfine interaction~\cite{DRGG}. It also behaves as
$\alpha_s^2N_c\sim 1/N_c$, however to fit the splitting numerically one needs to take $\alpha_s\approx 2$
whereas fits of deep inelastic scattering data and other phenomena tend to freeze $\alpha_s$ in the
infrared at the value about 0.5. Such value would give a tiny hyperfine splitting, hinting that it may
be irrelevant. The collective quantization interpretation is, numerically, more realistic.
Indeed, an estimate of the baryon moment of inertia is $I=mr^2\approx(1\,{\rm GeV})\cdot(0.5\,{\rm fm})^2$
yielding $1/I\approx 160\,{\rm MeV}$ as needed.

Another argument in favor of the $SU(2)_{{\rm iso}+{\rm space}}$ symmetry of the mean field comes
from the fact that baryons are strongly coupled to the pseudoscalar mesons ($g_{\pi NN}\approx 13$).
It means that there is a strong pseudoscalar field inside baryons; at large $N_c$ it is a classical mean field.
There is no way of writing down an {\it Ansatz} for the pseudoscalar field that would be odd with respect to space
inversion and simultaneously compatible with the $SU(3)_{\rm flav}\times SO(3)_{\rm space}$ symmetry. The minimal
extension of spherical symmetry is to write the ``hedgehog'' {\it Ansatz} ``marrying'' the isotopic and space
axes~\cite{footnote-2}:
\beq
\pi^a({\bf x})=\left\{\begin{array}{ccc}n^a\,F(r),& n^a=\frac{x^a}{r},& a=1,2,3,\\
0,&&a=4,5,6,7,8.\end{array}\right.
\la{hedgehog}\eeq

This {\it Ansatz} breaks the $SU(3)_{\rm flav}$ symmetry. If $m_s\!=\!m_u\!=\!m_d$, all flavor axes are
equivalent, therefore writing the pseudoscalar field in this form means nothing but naming the
$SU(3)$ axes. Analogously, the spontaneous magnetization in a ferromagnet can assume any direction,
and we can {\em name} the direction of magnetization as, say, the $z$ direction. If there is an external
magnetic field, even infinitesimal, it sets a preferred direction such that the spontaneous magnetization will be
along it. A nonzero magnetic field is analogous to the case of $m_s>m_u\!=\!m_d$ when the strange direction is privileged.
However, the {\it Ansatz} \ur{hedgehog} implies a spontaneous (as contrasted to explicit) symmetry breaking,
since $s$ quarks are treated in a totally different way than the $u,d$ ones, even if $m_s$ differs
infinitesimally from $m_u\!=\!m_d$.

Moreover, the {\it Ansatz} \ur{hedgehog} breaks the symmetry under independent space $SO(3)_{\rm space}$
and isospin $SU(2)_{\rm iso}$ rotations, and only a simultaneous rotation in both spaces leaves \ur{hedgehog}
invariant. Therefore, the {\it Ansatz} \ur{hedgehog} breaks spontaneously the original
$SU(3)_{\rm flav}\times SO(3)_{\rm space}$ symmetry down to the $SU(2)_{{\rm iso}+{\rm space}}$ symmetry.
This is precisely what is needed to obtain the correct baryon spectrum, where some excitations are large
(${\cal O}(1)$) and some are small (${\cal O}(1/N_c)$). We note that the splittings {\it inside} $SU(3)$
multiplets can be determined as a perturbation in $m_s$~\cite{Blotz}.

The full list of other possible mesonic fields in baryons (scalar, vector, axial, tensor), compatible with the
$SU(2)_{{\rm iso}+{\rm space}}$ symmetry is given in Ref.~\cite{D-09-1}.

\section{Baryons made of $u,d,s$ quarks}

Given the $SU(2)_{{\rm iso}+{\rm space}}$ symmetry of the mean field,
the Dirac Hamiltonian for quarks actually splits into two: one for $s$ quarks and the other
for $u,d$ quarks. It should be stressed that the energy levels for $u,d$ quarks on the one hand and for
$s$ quarks on the other are completely different, even in the chiral limit $m_s\to 0$.

The energy levels for $s$ quarks are classified by half-integer $J^P$ where
${\bf J}={\bf L}+{\bf S}$ is the angular momentum, and are $(2J+1)$-fold degenerate.
The energy levels for $u,d$ quarks are classified by integer $K^P$ where ${\bf K}={\bf T}+{\bf J}$
is the `grand spin' ($T$ is isospin), and are $(2K+1)$-fold degenerate.

All energy levels, both positive and negative, are probably discrete owing to confinement.
Indeed, a continuous spectrum would correspond to a situation when quarks are free at large
distances from the center, which contradicts confinement. One can model confinement {\it e.g.}
by forcing the effective quark masses to grow linearly at infinity.

According to the Dirac theory, all {\em negative}-energy levels, both for $s$ and $u,d$ quarks,
have to be fully occupied, corresponding to the vacuum. It means that there must be exactly
$N_c$ quarks antisymmetric in color occupying all degenerate levels with $J_3$ from $-J$ to $J$,
or $K_3$ from $-K$ to $K$; they form closed shells.
Filling in the lowest level with $E>0$ by $N_c$ quarks makes a baryon~\cite{DP-86,D-09}, see Fig.~1.
A similar picture arises in the chiral bag model~\cite{Hosaka}.

\begin{figure}[htb]
\begin{minipage}[]{0.45\textwidth}
\vspace{-4.5cm}
\includegraphics[width=\textwidth]{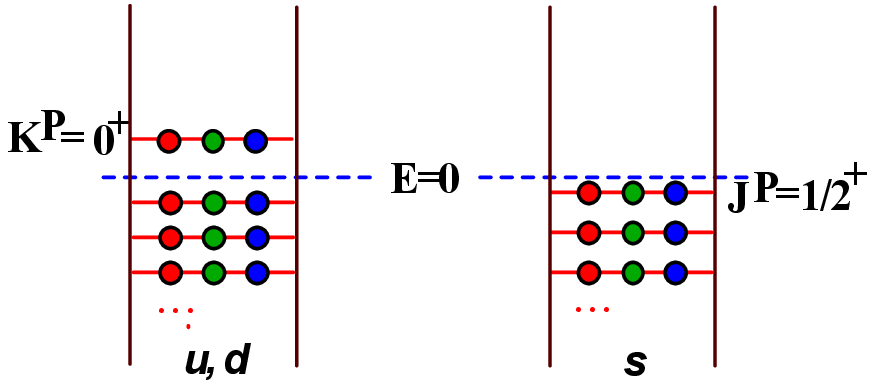}
\end{minipage}
\hspace{1.0cm}
\begin{minipage}[t]{0.45\textwidth}
\includegraphics[width=\textwidth]{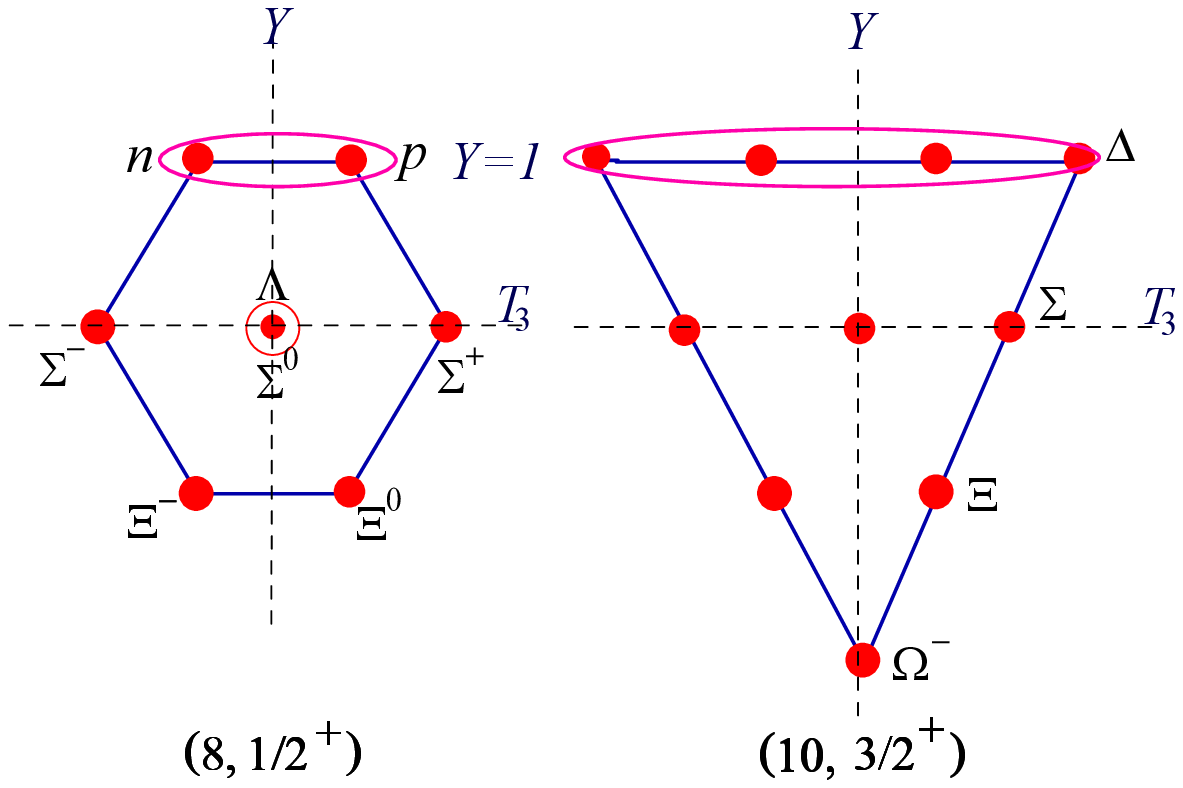}
\end{minipage}
\caption{Filling $u,d$ and $s$ shells for the ground-state baryon (left), and the two lowest
baryon multiplets that follow from quantizing the rotations of this filling scheme (right). }
\label{fig:1}
\end{figure}

The mass of a baryon is the aggregate energy of all filled states, and being a functional
of the mesonic field, it is proportional to $N_c$ since all quark levels are degenerate in color.
Therefore quantum fluctuations of mesonic field in baryons are suppressed as $1/N_c$ so that
the mean field is indeed justified.

Quantum numbers of the lightest baryons are determined from the quantization of the rotations
of the mean field, leading to specific $SU(3)$ multiplets that reduce at $N_c\!=\!3$ to the octet
with spin $\half$ and the decuplet with spin $\frac{3}{2}\;$, see {\it e.g.}~\cite{DP-08} and the Appendix.
Witten's quantization condition $Y'\!=\!\frac{N_c}{3}$~\cite{Witten-Sk} follows trivially from
the fact that there are $N_c$ $\;u,d$ valence quarks each with the hypercharge $\frac{1}{3}$~\cite{Blotz}.
Therefore, the ground state shown in Fig.~1 entails in fact 56 rotational states.
The splitting between the centers of the multiplets $({\bf 8},\half^+)$ and $({\bf 10}, \frac{3}{2}^+)$
is ${\cal O}(1/N_c)$, and the splittings inside multiplets can be determined as a perturbation
in $m_s$~\cite{Blotz}.


The lowest baryon resonance beyond the rotational excitations of the ground state
is the singlet $\Lambda(1405,\half^-)$. Apparently, it can be obtained only as an excitation
of the $s$ quark, and its quantum numbers must be $J^P=\half^-$~\cite{D-09}, see transition
{\it 1} in Fig.~2.

The existence of an $\half^-$ level for $s$ quarks automatically implies that there is a
particle-hole excitation of this level by an $s$ quark from the $\half^+$ level.
I identify this transition {\it 2} with $N(1535,\half^-)$~\cite{D-09}. At $N_c\!=\!3$ it is
predominantly a pentaquark state $u(d)uds\bar s$ (although it has also a nonzero three-quark
Fock component). This explains its large branching ratio in the $\eta N$ decay~\cite{Zou},
a long-time mystery. We also see that, since the highest filled level for $s$ quarks is lower
than the highest filled level for $u,d$ quarks, $N(1535,\half^-)$ must be {\em heavier} than
$\Lambda(1405,\half^-)$: the opposite prediction of the non-relativistic quark model has been
always of some concern. Subtracting $1535-1405=130$, I find that the $\half^+$ $s$-quark level
is approximately 130 MeV lower in energy than the valence $0^+$ level for $u,d$ quarks. This is
an important number which will be used below. The transition entails its own rotational band
discussed in the Appendix.

\begin{figure}[htb]
\begin{minipage}[]{.5\textwidth}
\includegraphics[width=\textwidth]{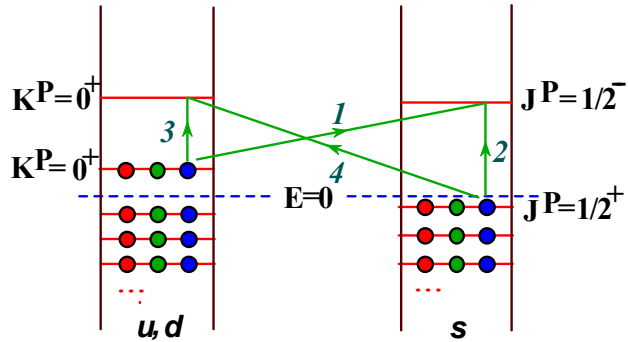}
\end{minipage}
\caption{The existence of the two lowest excited levels -- one for the $u,d$ quarks and another for the
$s$ quarks -- implies four resonances shown by arrows.
The transitions correspond to:
{\it 1}: $\Lambda(1405,{1/2}^-)$,
{\it 2}: $N(1535,{1/2}^-)$,
{\it 3}: $\!\!N(1440,{1/2}^+\!)\!$,
{\it 4}: $\!\!\Theta^+(1530,{1/2}^+\!)\!$.
Each transition generally entails its own rotational band of $SU(3)$ multiplets.}
\label{fig:2}
\end{figure}

The low-lying Roper resonance $N(1440,\half^+)$ requires an excited one-particle $u,d$ state
with $K^P=0^+$ (or $1^+$)~\cite{D-09}, see transition {\it 3}. Just as the ground state nucleon, it is part
of the excited $({\bf 8'},\half^+)$ and $({\bf 10'}, \frac{3}{2}^+)$ split as $1/N_c$.
Such identification of the Roper resonance solves another problem of the non-relativistic
model where $N(1440,\half^+)$ must be heavier than $N(1535,\half^-)$. In our approach they are unrelated.

Given that there is an excited $0^+$ level for $u,d$ quarks, one can put there a quark taking it as well
from the $s$-quark $\half^+$ shell, see transition {\it 4}.
It is a particle-hole excitation with the valence $u,d$ level left untouched, its quantum numbers being
$S=+1,\;T=0,\;J^P=\half^+$. At $N_c\!=\!3$ it is a pentaquark state $uudd\bar s$, precisely
the exotic $\Theta^+$ baryon predicted in Ref.~\cite{DPP-97} from related but somewhat different considerations.
The quantization of its rotations produces the antidecuplet $({\bf \overline{10}},\half^+)$.
In our original prediction the ${\cal O}(1)$ gap between $\Theta^+$ and the nucleon was due to the
rotational energy only, whereas here the main ${\cal O}(1)$ part of that gap is due to the one-particle
levels, while the rotational energy is ${\cal O}(1/N_c)$, see the Appendix.
Methodologically, it is now more satisfactory.

In nuclear physics, excitations generated by the axial current $j_{\mu\,5}^\pm$,
when a neutron from the last occupied shell is sent to an unoccupied proton level or {\it v.v.}
are known as Gamov--Teller transitions~\cite{BM}. Thus our interpretation of the \Th is that it is a
Gamov--Teller-type resonance long known in nuclear physics.

An unambiguous feature of our picture is that {\bf the exotic pentaquark \Th is a consequence of the existence
of three well-known resonances and must be light.} Indeed, the $\Theta^+$ mass can be estimated from the apparent
sum rule following from Fig.~2~\cite{D-09}: $m_{\Theta}\approx 1440+1535-1405\approx 1570\,{\rm MeV}$.
Since the $N(1440)$ and $N(1535)$ resonances are broad such that their masses are not well defined,
there is a numerical uncertainty in this equation. For example, if one uses the pole positions of the
resonances the equation reads $m_{\Theta}\approx 1365+1510-1405\approx 1470\,{\rm MeV}$. Therefore, it is
fair to say that the sum rule predicts $m_\Theta=1520\pm 50\,{\rm MeV}$. This is in remarkable agreement
with the claimed masses of the \Th: $m_\Theta= 1524\pm 2 \pm 3\,{\rm MeV}$~\cite{LEPS-2},
$1537\pm 2\,{\rm MeV}$~\cite{DIANA-2}, $1523\pm 2\pm 3\,{\rm MeV}$~\cite{SVD-2},
$1521.5\pm 1.5\pm 2.8/1.7\,{\rm MeV}$~\cite{ZEUS}, $1528\pm 2.6\pm 2.1\,{\rm MeV}$~\cite{HERMES}.
For a possible explanation why \Th is seen in some experiments while not observed in other see Ref.~\cite{Azimov}.

To account for higher baryon resonances one has to assume that there are higher one-particle levels,
both in the $u,d$- and $s$-quark sectors, to be published elsewhere~\cite{D-09-1}.

\section{Baryon resonances from rotational bands}

A filling scheme of one-particle quark levels by itself does not tell us what are the quantum numbers
of the state. The filling scheme treats $u,d$ quarks and $s$ quarks differently and therefore violates the
$SU(3)_{\rm flav}$ and also $SO(3)_{\rm space}$ symmetries. Only the $SU(2)_{{\rm iso}+{\rm space}}$ symmetry
of simultaneous isospin and compensating space rotations is preserved. In the chiral limit (which I assume
for the time being) an arbitrary $SU(3)_{\rm flav}$ rotation of the mean field and hence of what we call $u,d,s$ quarks
does not change the energy of the state. The same is true for the $SO(3)_{\rm space}$ rotation. However, if
$SU(3)_{\rm flav}$ and $SO(3)_{\rm space}$ rotations are slowly dependent on time, they generate a shift in
the energy of the system; it is called the rotational energy. Being quantized according to the general
quantization rules for rotations, it produces states with definite $SU(3)_{\rm flav}$ quantum numbers and spin.

Thus the original $SU(3)_{\rm flav}\times SO(3)_{\rm space}$ symmetry broken spontaneously by a `hedgehog'
{\it Ansatz} of the mean field, is restored when flavor and space rotations are accounted for.
Each transition in Fig.~2 generally entails ``rotational bands'' of $SU(3)$ multiplets with definite spin
and parity. The short recipe of getting them is: Find the hypercharge $Y'$ of the given excitation from the
number of $u,d,s$ quarks involved; only those multiplets are allowed that contain this $Y'$.
Take an allowed multiplet and read off the isospin(s) $T'$ of particles at this value of $Y'$.
The allowed spin of the multiplet obeys the angular momentum addition law:
\beq
{\bf J}={\bf T'}+{\bf J_1}+{\bf J_2}
\la{Tprime}\eeq
where $J_{1,2}$ are the initial and final momenta of the $s$ shells involved in the
transition. (If nonzero $K$ shell is involved in the transition the quantization rule is more complex.)
The mass of the center of an allowed rotational multiplet does not depend on ${\bf J}$
but only on ${\bf T'}$ according to the relation~\cite{DP-04}
\beq
{\cal M}={\cal M}_0+\frac{C_2(p,q)-T'(T'+1)-\frac{3}{4}{Y'}^2}{2I_2}+\frac{T'(T'+1)}{2I_1}
\la{Erot}\eeq
where $C_2(p,q)=\third(p^2+q^2+pq)+p+q$ is the quadratic Casimir eigenvalue of the $SU(3)$ multiplet characterized
by $(p,q)$, $I_{1,2}={\cal O}(N_c)$ are moments of inertia. After the rotational band for a given transition is constructed,
one has to check if the rotational energy of a particular multiplet is ${\cal O}(1/N_c)$ and not ${\cal O}(1)$,
and if it is compatible with Fermi statistics at $N_c\!=\!3$: some {\it a priori} possible multiplets drop out.
One gets a satisfactory description of all light baryon resonances up to about 2 GeV, to be published separately~\cite{D-09-1}.

\section{Charmed and bottom baryons, the lowest multiplets}

If one of the light quarks in a light baryon is replaced by a heavy $b$ or $c$ quark,
there are still $N_c\!-\!1$ light quarks left. At large $N_c$, they form {\em the same} mean field as
in light baryons, with the same sequence of Dirac levels, up to $1/N_c$ corrections. The heavy quark
contributes to the mean $SU(3)_{\rm flav}$ symmetric field but it is a $1/N_c$ correction, too. It means
that at large $N_c$ one can {\em predict} the spectrum of the $Qq\ldots q$ (and $Qq\ldots qq\bar q$) baryons from
the spectrum of light baryons. At $N_c\!=\!3$ one does not expect qualitative difference with the $N_c\to\infty$ limit,
although $1/N_c$ corrections should be kept in mind. I consider the heavy quark as a non-relativistic particle
having spin $J_h=\half$. $SU(4)_{\rm flav}$ symmetry is badly violated and is of no guidance.

The filling of Dirac levels for the ground-state $c$ (or $b$) baryon is shown in Fig.~3, left:
there is a hole in the $0^+$ shell for $u,d$ quarks as there are only $N_c-1$ quarks there, in an antisymmetric
state in color. Adding the heavy quark makes the full state `colorless'.

\begin{figure}[htb]
\begin{minipage}[b]{.95\textwidth}
\includegraphics[width=0.45\textwidth]{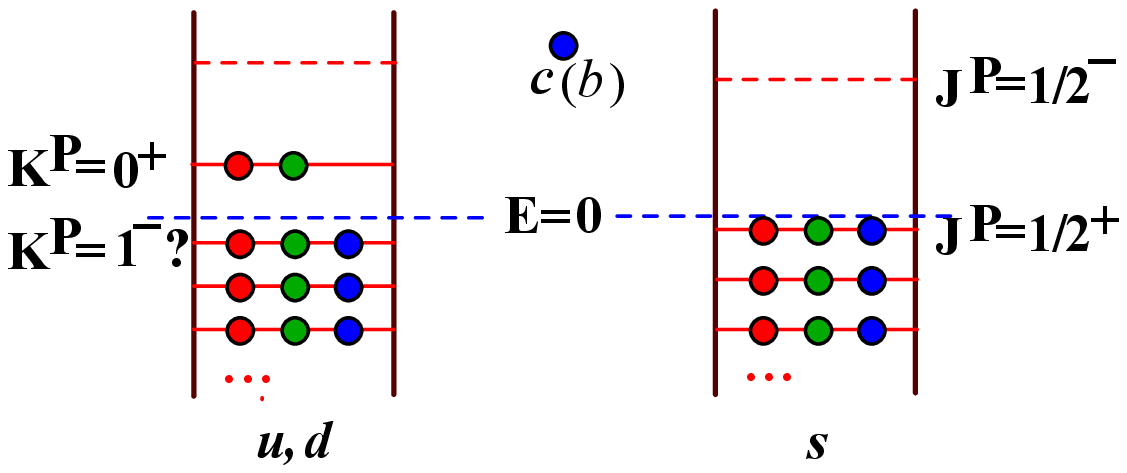}\hspace{1.2cm}
\includegraphics[width=0.45\textwidth]{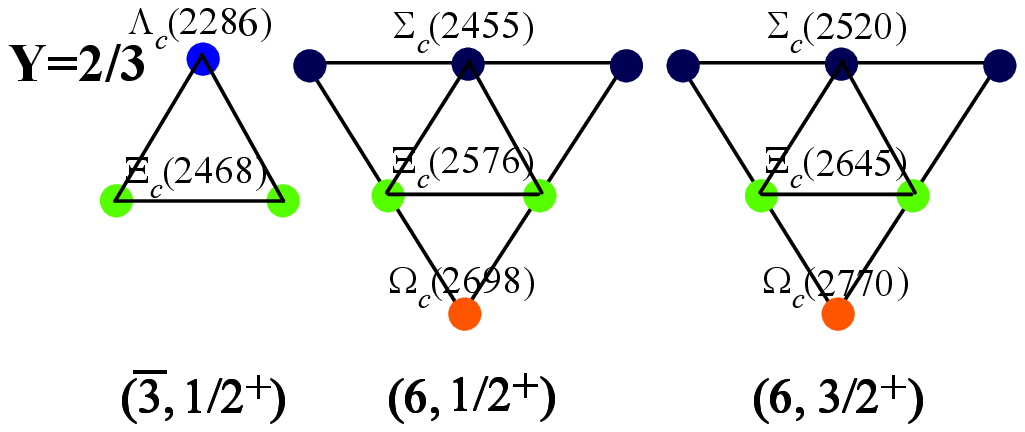}
\end{minipage}
\caption{Filling $u,d$ and $s$ shells for the ground-state charmed baryons (left), and $SU(3)$ multiplets
generated by this filling scheme (right): $({\bf \bar 3},{1/2}^+)$, $({\bf 6},{1/2}^+)$ and $({\bf 6},{3/2}^+)$.}
\label{fig:3}
\end{figure}

As in the case of light baryons, the filling scheme by itself does not tell us what are the quantum numbers
of the state: they arise from quantizing the $SU(3)_{\rm flav}$ and $SO(3)_{\rm space}$ rotations of the given
filling scheme. Let us do it for the ground-state baryons.

First of all, we determine the hypercharge of the filling scheme: in this case it is $Y'=\frac{1}{3}(N_c\!-\!1)$
since there are $N_c\!-\!1$ $u,d$ quarks each having hypercharge one third. At $N_c=3$ one has $Y'=\frac{2}{3}$.
There are two $SU(3)$ multiplets containing particles with hypercharge $\frac{2}{3}$: the anti-triplet ${\bf \bar 3}$
$(p\!=\!0,q\!=\!1)$ and the sextet ${\bf 6}$ $(p\!=\!2,q\!=\!0)$, therefore these are the allowed multiplets,
see Fig.~3, right. What are their spins?

In the ${\bf\bar 3}$ representation, there is one particle with $Y'=\frac{2}{3}$ hence its isospin $T'=0$.
The possible spin of the multiplet is found from \Eq{Tprime} which needs to be modified to include the
spin of the heavy quark $J_h$:
\beq
{\bf J}={\bf T'}+{\bf J_1}+{\bf J_2}+{\bf J_h}.
\la{Tprimeh}\eeq
In this case $J_1=J_2=0$ since $s$ quarks are not involved, $T'=0$, and $J_h=\half$. Therefore, the only
possible spin of the anti-triplet is $\half$, and parity plus. Its rotational energy is, according to \Eq{Erot},
\beq
E_{\rm rot}^{({\bf\bar 3})}=\frac{1}{2I_2}.
\la{Erotanti3}\eeq

In the ${\bf 6}$ representation, there are three particles with $Y'=\frac{2}{3}$ hence their isospin $T'=1$.
From \Eq{Tprimeh} one finds then that there are {\it two} sextets, one with spin $\half$ and another with
spin $\frac{3}{2}$. They are degenerate in the leading order as the rotational energy \ur{Erot} depends only
on $T'$ but not on the spin:
\beq
E_{\rm rot}^{({\bf 6})}=\frac{1}{2I_2}+\frac{1}{I_1}.
\la{Erot6}\eeq

Thus the filling scheme in Fig.~3, left, implies three $SU(3)$ multiplets: $({\bf \bar 3}, \left.\half\right.^+),\;
({\bf 6}, \left.\half\right.^+)$ and $({\bf 6}, \left.\frac{3}{2}\right.^+)$, see Fig.~3, right. The last two
are degenerate (but the degeneracy is lifted in the next $1/N_c^2$ order and also from the $1/m_h$ corrections)
whereas the center of the anti-triplet is separated from the center of the sextets by the rotational energy
$\Delta E_{\rm rot} =\frac{1}{I_1}$. The splitting {\em inside} multiplets owing to the explicit violation of
$SU(3)$ by the strange quark mass is ${\cal O}(m_sN_c)$. If $m_s$ is treated as a small perturbation,
$m_s={\cal O}(1/N_c^2)$, as I claim it should~\cite{D-09}, the splitting inside the sextet must be equidistant
to a good accuracy. Let us confront these predictions with current data.

There are good candidates for the above ground-state multiplets: $\Lambda_c(2286)$ and $\Xi_c(2468)$ for
$({\bf \bar 3},{1/2}^+)$; $\Sigma_c(2455)$, $\Xi_c(2576)$ and $\Omega_c(2698)$ for $({\bf 6},{1/2}^+)$;
finally $\Sigma_c(2520)$, $\Xi_c(2645)$ and $\Omega_c(2770)$ presumably form $({\bf 6},{3/2}^+)$, see Fig.~3, right.
Strictly speaking the $J^P$ quantum numbers of most of these baryons are not measured directly but there
is not much doubt they differ from the above assignments. Assuming they are correct, the observed
parity-plus charmed baryons form precisely those multiplets that follow from the collective quantization.

The splittings inside the two sextets are equidistant to high accuracy, confirming that $m_s$ can be treated
as a small perturbation. Were $m_s$ ``not small'', there would be substantial ${\cal O}(m_s^2)$ corrections
to the masses, which would violate the equidistant character of the sextets spectrum.

The centers of the three multiplets are at
\bea\n
m\left({\bf \bar 3},\left.1/2\right.^+\right)&=&\frac{2287+2*2468}{3}=2408\;{\rm MeV},\\
\n
m\left({\bf 6},\left.1/2\right.^+\right)&=&\frac{3*2455+2*2576+2698}{6}=2536\;{\rm MeV},\\
m\left({\bf 6},\left.3/2\right.^+\right)&=&\frac{3*2520+2*2645+2770}{6}=2603\;{\rm MeV}.
\la{m366}\eea
Although the two sextets are not exactly degenerate, their splitting 67 MeV (an unaccounted $1/N_c^2$ effect)
is much less than the splitting between the anti-triplet and the mean mass of the sextets, which is
\beq
\frac{2536+2603}{2}-2408=162\;{\rm MeV}=E_{\rm rot}^{({\bf 6})}-E_{\rm rot}^{({\bf\bar 3})}=\frac{1}{I_1}={\cal O}(1/N_c).
\la{DeltaE63}\eeq
Furthermore, this number should be compared with the moment of inertia following from the splitting between
{\em light} baryons, $({\bf 10},\left.\frac{3}{2}\right.^+)$ and $({\bf 8},\left.\frac{1}{2}\right.^+)$, yielding
$1/I_1=153\,{\rm MeV}$, see Section II. The proximity of the two completely different determinations of the
moment of inertia supports the basic idea that it is reasonable to view both light and heavy baryons from the
same large-$N_c$ perspective~\cite{footnote-3}.

\section{Charmed and bottom baryons, excited states}

There are higher $SU(3)$ multiplets with $Y'=\frac{2}{3}$, however a closer inspection shows that the corresponding
rotational excitations have large ${\cal O}(1)$ energies and not ${\cal O}(1/N_c)$ as requested for the rotational
states. Therefore, those higher rotational states are, strictly speaking, beyond control. Higher parity-plus
heavy baryon resonances should arise as one-particle and particle-hole excitations, like for light baryons.

As to parity-minus states, there are several possibilities to construct them. The first is to excite one of the
$u,d$ quarks from the $0^+$ valence level to the first excited level for $s$ quarks, which is $\half^-$, see
Fig.~2, transition {\it 1}. It would be then an analogue of $\Lambda(1405,\half^-)$. There is, however, an
argument against it. The transition has $Y'=-\frac{1}{3}$ which is possible with the ${\bf\bar 3}$ and ${\bf 6}$
representations but now it corresponds to $T'=\half$ in both cases. This difference with ground-state multiplets
has a dramatic consequence: both multiplets have an ${\cal O}(1)$ rotational energy and hence should be discarded.
[To check the analytical $N_c$ dependence one has to construct the prototype multiplets that reduce at $N_c=3$
to those under consideration, see the Appendix.]

The second possible way of making parity-minus states is to excite the $s$ quark from the highest filled $\half^+$
level to the $\half^-$ level. It would be an analogue of $N(1535,\half^-)$, see Fig.~2, transition {\it 2}. The
quantization of rotations about this excitation along the lines presented above, leads to three degenerate anti-triplets
$2\times({\bf\bar 3},1/2^-),\;({\bf\bar 3},3/2^-)$ and many sextets with various spins. However, this interpretation
is not too realistic either. First, there are only two observed anti-triplets presumably with negative parity,
$\Lambda_c(2595,{1/2}^-),\;\Xi_c(2790,{1/2}^-?)$ and $\Lambda_c(2625,{3/2}^-?),\;\Xi_c(2815,{3/2}^-?)$. There are no
candidates for the second $({\bf\bar 3}, 1/2^-)$ although it may be found in future. Second, the average mass
of the two $\Lambda_c$'s is only 300 MeV higher than the lowest charmed baryon $\Lambda_c(2786.5, {1/2}^+)$, whereas
from light baryons we expect that this excitation energy is about $1535-940\approx 600\;{\rm MeV}$. It contradicts
the basic large-$N_c$ concept that the one-particle levels for heavy baryons do not differ much from those for light
baryons.

Therefore at the moment I think that the most plausible construction of the lowest parity-minus states is to assume
that there is a $1^-$ level for $u,d$ quarks {\em below} the valence $0^+$ level, see Fig.~3, left. It has to be
filled in in the ground state as it has $E<0$ but since there is a hole in the valence $0^+$ level there is
an excitation when one $u,d$ quark from the $1^-$ level fills the hole in the $0^+$ shell. Such excitation is absent
for light baryons (since the valence $0^+$ shell is fully filled there), therefore no previous knowledge prevents
us from assuming that the excitation energy is only 300 MeV.

The quantization of rotations about such excitation produces two degenerate anti-triplets
$({\bf\bar 3},1/2^-)$ and $({\bf\bar 3},3/2^-)$, exactly as needed for phenomenology. However, this transition
generates also a number of higher mass almost degenerate sextets with spin from $1/2$ to $5/2$, none of which
has been observed so far. Unfortunately, experimental knowledge of the parity-minus heavy baryons is too scarce
to choose between different interpretations.

\section{Charmed and bottom baryons, exotic states}

Our new observation is that there is a Gamov--Teller-type transition when the axial current annihilates a
strange quark in the $\half^+$ shell, and creates an $u$ or $d$ quark in the $0^+$ shell (see Fig.~4, left),
like in the case of the $\Theta^+$. In heavy baryons it is even more trivial as there is a hole in the $0^+$
valence shell from the start. Filling in this hole means making charmed (or bottom) pentaquarks which I name
``Beta baryons''~\cite{footnote-4}, ${\cal B}_c^{++}=cuud\bar s,\;{\cal B}_c^+=cudd\bar s$,
and ${\cal B}_b^+=buud\bar s,\;{\cal B}_b^0=budd\bar s$. While the existence of \Th requires an excited (`Roper')
one-particle level, the existence of the ${\cal B}_{c,b}$ baryons needs only the ground-state level which is
undoubtedly there. In this sense, the ${\cal B}_{c,b}$ baryons are more basic than the \Th.

What are the $SU(3)$ multiplets corresponding to this excitation? The hypercharge is
$Y'=3*\frac{1}{3}-(-\frac{2}{3})=\frac{5}{3}$. The lowest $SU(3)$ representation containing particles with
$Y'=\frac{5}{3}$ is the {\em anti-decapenta} (${\bf\overline{15}}$)-plet ($p=1,q=2$)~\cite{footnote-5}, see Fig.~4, right.
Therefore, this is an allowed multiplet generated by the transition. There are two particles with $Y'=\frac{5}{3}$,
hence their isospin is $T'=\half$. The allowed spin is given by \Eq{Tprimeh} where one puts $J_1=\half,\;J_2=0$
and obtains that the possible spins of the multiplets are $\half$ (twice) and $\frac{3}{2}$, parity plus.
All of them are degenerate in the leading order in $1/N_c$ but split in the next-to-leading order.

Thus the Gamov--Teller-type transition shown in Fig.~4, left, induces three almost degenerate multiplets:
$2\times ({\bf\overline{15}},\left.1/2\right.^+)$ and $({\bf\overline{15}}, \left.3/2\right.^+)$.

\begin{figure}[htb]
\begin{minipage}[t]{.95\textwidth}
\includegraphics[width=0.52\textwidth]{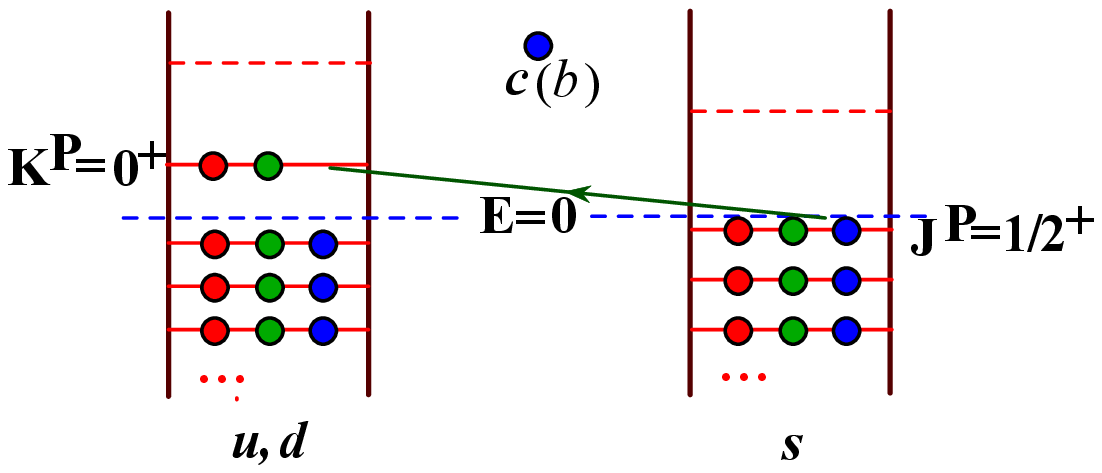}\hspace{1.2cm}
\includegraphics[width=0.37\textwidth]{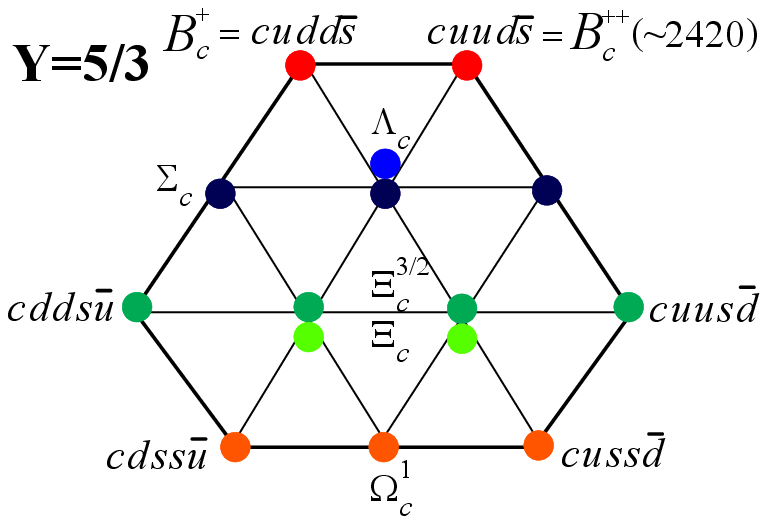}
\end{minipage}
\caption{The arrow shows the lowest Gamov--Teller excitation (left) leading to charmed
pentaquarks forming ${\bf\overline{15}}$ (right).}
\label{fig:4}
\end{figure}

The six baryons at the corners of the hexagon in Fig.~4, right, are explicitly exotic: their quantum numbers cannot
be achieved from 3-quark states. The rest 9 baryons are crypto-exotic: they are mainly pentaquarks but have the quantum
numbers of the ground-state baryons belonging to ${\bf\bar 3}$ and ${\bf 6}$ representations, and can mix with them.
The mixing is an $SU(3)$ violating effect, the mixing angle being $\theta={\cal O}(m_sN_c^2/\Lambda)$ where
$\Lambda\sim 1\,{\rm GeV}$ is a typical scale in strong interactions. Actually the isotopic quadruplet $\Xi_c^{3/2}$
and the triplet $\Omega_c^1$ mix up with the corresponding members of the ${\bf\bar 3}$ and ${\bf 6}$ only through
isospin breaking, therefore this mixing can be neglected. The mixing of $\Lambda_c$, $\Sigma_c$ and $\Xi_c$ leads to
a shift in the physical baryon masses, that is quadratic in $m_s$; it is of the order of $m_s^2N_c^3/\Lambda$.
The fact that baryons in the sextets are almost equidistant means that in practice the mixing is numerically small.
Probably more important is the mixing between the two $({\bf\overline{15}},{1/2}^+)$-plets with identical quantum
numbers: one goes up, the other goes down.

\section{Mass splitting in the exotic anti-decapenta-plet}

The ${\bf\overline{15}}$-plets are split in the leading ${\cal O}(m_sN_c)$ order, as usually. The $SU(3)$-violating
term in the QCD Lagrangian, $m_s\bar s s$, has a singlet and an octet pieces; it is the octet piece that
leads to the splitting. Since there are two ways of getting an octet in the direct product of
${\bf\overline{15}}\otimes{\bf 15}$ there are two mass parameters $m_{1,2}$ that determine
all mass shifts in the ${\bf\overline{15}}$. [This is similar to the octet baryons where there are also two parameters,
and distinct from the sextet, decuplet and antidecuplet where there is only one, and consequently the masses
are equidistant.] In what follows I ignore the next-to-leading mixing of the ${\bf\overline{15}}$-plets with the
${\bf\bar 3}$ and ${\bf 6}$, and of the two $({\bf\overline{15}},{1/2}^+)$ between themselves.

A simple exercise in the $SU(3)$ algebra (which I suppress) leads to the following masses of the members of
the ${\bf\overline{15}}$-plet, for each horizontal line in Fig.~4, right, from top to bottom:
\bea\la{masses15}
{\cal B}_c^{++,+}:\quad M_1&=&M_0-m_1-m_2,\\
\n
\Sigma_c^{++,+,0}:\quad M_2&=&M_0-m_1-\frac{m_2}{4},\\
\n
\Lambda_c^+:\quad M_3&=&M_0+\frac{m_1}{2}-\frac{5m_2}{8},\\
\n
\Xi_c^{++,+,0,-}:\quad M_4&=&M_0-m_1+\frac{m_2}{2},\\
\n
\Xi_c^{+,0}:\quad M_5&=&M_0+\frac{5m_1}{4}-\frac{m_2}{16},\\
\n
\Omega_c^{+,0,-}:\quad M_6&=&M_0+2m_1+\frac{m_2}{2},
\eea
where
$$
M_0=\frac{1}{15}(2M_1+3M_2+M_3+4M_4+2M_5+3M_6)
$$
is the center of the ${\bf\overline{15}}$-plet. There are 6 different masses $M_{1\!-\!6}$ expressed through 3
parameters, therefore there are 3 relations, analogous to the Gell-Mann-Okubo relation for the octet:
\bea\n
M_1+M_4&=&2M_2,\\
\n
5M_1+2M_2+2M_5&=&9M_3,\\
\n
4M_1+M_2+M_6&=&6M_3.
\eea
These combinations do not depend on the $SU(3)$-violating parameters $m_{1,2}$. However, in the large-$N_c$
approach $m_{1,2}$ are related to the splittings inside all other multiplets: ${\bf\bar 3},\;{\bf 6},\;
{\bf 8},\;{\bf 10}$ and ${\bf\overline{10}}$. These relations will be considered elsewhere.

We see that the lightest is the exotic doublet ${\cal B}_c$, and the heaviest is the exotic triplet $\Omega_c^1$.
Since we know the separation between the ${1/2}^+$ level for $s$ quarks and the $0^+$ level for
$u,d$ quarks from fitting the light baryon resonances (it is 130 MeV, see Section III), and assuming that it
does not change for heavy baryons (as it would be at $N_c\to\infty$), we estimate the mass of the ${\cal B}_c^{++,+}$
pentaquarks at about $m(\Lambda_c)+130\,{\rm MeV}=2420\,{\rm MeV}$. The corresponding bottom pentaquarks ${\cal B}_b^{+,0}$
mass is about $m(\Lambda_b)+130\,{\rm MeV}=5750\,{\rm MeV}$. These are very light masses.

The accuracy of this prediction is ${\cal O}(1/N_c)\sim 150\,{\rm MeV}$ but there is a 360 MeV margin
below the threshold for strong decays ${\cal B}_c\to\Lambda_cK$ (2780 MeV), ${\cal B}_b\to\Lambda_bK$ (6110 MeV).
It means that such light charmed and bottom baryons have no strong decays, which makes their observation feasible.

Using the rule of thumb that each lower line in the $SU(3)$ weight diagram is approximately 140 MeV
heavier than the previous, the expected mass of another exotic pentaquark, the quadruplet $\Xi_c^{3/2}$,
is about 2700 MeV. However, this is above the threshold for the strong decay, for example
$\Xi_c^-(csdd\bar u)\to\Xi_c^0(csd)+\pi^-(d\bar u)$, at 2610 MeV. Therefore, the exotic pentaquarks $\Xi_c^{3/2}$
can manifest themselves as extremely narrow peaks in $\Xi_c\pi$ mass distributions. The same applies
to the exotic $\Omega_c^1(\sim 2850)$ decaying into $\Omega_c\pi$ (threshold at 2840 MeV).

Charmed pentaquarks have been considered by Wu and Ma in another approach~\cite{WuMa}; however, these
authors get far larger masses and in addition pentaquarks with $\bar c$ quarks appear almost degenerate
with those made of $c$ quarks. This is not the case in the present scheme where {\em anti}-charmed pentaquarks
are ${\cal O}(1)$, that is substantially, heavier than the charmed ones.

How to make a $qqqq\bar Q$ pentaquark in the present approach? Apparently the $0^+$ shell for $u,d$ quarks must be
completed, and one has to put somewhere the fourth quark to make the state `colorless'. The first two excited
states, the $\half^-$ level for $s$ quarks and the excited (`Roper') $0^+$ level for $u,d$ quarks, are rather
close in light baryons, see Fig.~2. They may be reshuffled somewhat by $1/N_c$ corrections as one goes from
light to heavy baryons, therefore which one is lower in heavy baryons is not clear beforehand. Assuming it
is the $\half^-$ level, the lowest anti-charmed pentaquarks are $P_{\bar cs}=uuds\bar c,\;udds\bar c$ of
Gignoux {\it et al.}~\cite{Gignoux} and Lipkin~\cite{Lipkin}; they belong to the ${\bf 3}$ representation
and have negative parity. The allowed spins are $\half$ (twice) and $\frac{3}{2}$, according to
the quantization rules formulated in Sections IV and V, see Fig.~5.

\begin{figure}[htb]
\begin{minipage}[]{.7\textwidth}
\includegraphics[width=\textwidth]{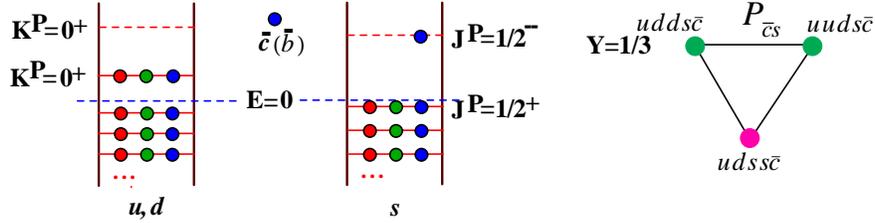}
\end{minipage}
\caption{{\em Anti}-charmed (anti-bottom) pentaquarks with negative parity; filling scheme for four quarks (left)
and the multiplets it generates (right): $2\times ({\bf 3},{1/2}^-)$ and $({\bf 3},{3/2}^-)$.}
\label{fig:5}
\end{figure}

Assuming the lowest excited state is the $0^+$ level, the lowest anti-charmed pentaquark is
$\Theta_c=uudd\bar c$ of Karliner and Lipkin~\cite{KarLip} belonging to the ${\bf\bar 6}$ representation;
it has then spin-parity $\half^+$, see Fig.~6. Such $\Theta_c$ is a direct analog of the light-quark \Th
as it also arises from exciting the `Roper' level.

\begin{figure}[htb]
\begin{minipage}[]{.7\textwidth}
\includegraphics[width=\textwidth]{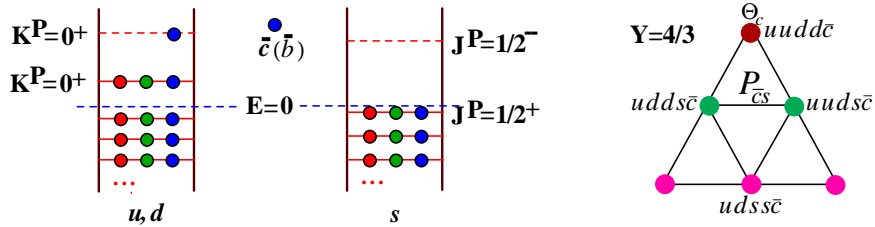}
\end{minipage}
\caption{{\em Anti}-charmed (anti-bottom) pentaquarks with positive parity; filling scheme for four quarks (left)
and the multiplet it generates (right): $({\bf\bar 6},{1/2}^+)$.}
\label{fig:6}
\end{figure}

To estimate the mass of anti-charmed pentaquarks, we assume that the valence $0^+$ level for $u,d$ quarks
is at about 100 MeV. It must be positive otherwise it would belong to the vacuum state, but less than 130 MeV
otherwise the $\half^+$ level for $s$ quarks would have positive energy. From light baryons we know that
the two excited levels are about 460 MeV higher than the valence $u,d$ shell. Therefore the lightest anti-charmed
pentaquarks are expected at about $m(\Lambda_c)+100+460\approx 2850\,{\rm MeV}$. This is slightly above the strong-decay
threshold for the $\Theta_c\to D^-p$ (2810 MeV) and slightly below the threshold for $P_{\bar cs}\to D_sp$ (2910 MeV).
The ${\cal O}(1/N_c)\sim 150\,{\rm MeV}$ precision in our mass predictions does not allow at present a definite
conclusion whether the anti-charmed baryons are stable against strong decays, which is critical for their
observation. The important point is that anti-charmed (anti-bottom) pentaquarks are essentially
($\sim 500\,{\rm MeV}$) heavier than the charmed (bottom) pentaquarks. Allowing even for a 360 MeV uncertainty
in numerics, Beta baryons ${\cal B}_{b,c}$ remain below the threshold for strong decays!

\section{Production rate of Beta baryons, and decay signatures}

In principle, ${\cal B}_{b,c}$ baryons can be produced whenever charm (bottom) is produced. However, the production
rate is expected to be very low. It is affected by the general suppression of charm (bottom) production, and by the
small coalescence factor specific for the production of objects built of many constituents. Therefore, high-energy,
high-luminosity machines like LHC have better chances.

It is very difficult to make a reliable estimate of the production rate, say, at LHC, therefore I make a pessimistic
estimate~\cite{footnote-6}. The number of charmed baryons produced in the central rapidity range (where it is maximal)
is estimated as $dN/dy \sim 10^{-3}$. For bottom quarks it is several times less. The number of anti-deuterons
produced at LHC is expected at the level of $dN/dy\sim 10^{-4}$. Deuterons are 6 quarks so the rate gives an idea
of the coalescence factor for a 5-quark system, too. To get the lower bound for the production rate for the pentaquark
${\cal B}_c$ baryons I am inclined to multiply the two probabilities and obtain for the LHC
\beq
\frac{dN^{{\cal B}_c}}{dy}\sim 10^{-7},\qquad y\approx 0.
\la{rate}\eeq
This is low enough but one looses even more when a specific channel is chosen to trigger the decay of ${\cal B}_c$.
From the experience with `ordinary' charmed baryons we know that there are very many decay channels, the largest
branching ratios being at the level of 1\%. Therefore, it is important to choose a decay channel with as low background
as possible, rather than seeking for a dominant decay mode. ${\cal B}_c^{++}$ has a remarkable decay into $p\pi^+$
proceeding through the Cabibbo-unsuppressed annihilation $c\bar s\to u\bar d$. However, this decay has probably
a large background even if events are selected with protons spatially displaced from the reaction vertex. I expect
that the ${\cal B}_c$ lifetime is of the same order as that of normal charmed baryons, {\it i.e.} $10^{-13}\,s$,
meaning that its decay can be resolved in a vertex detector. In addition, the in-flight Cabibbo-unsuppressed decay
$c\to su\bar d$ is probably faster than annihilation.

The ${\cal B}_c^{++}\to\bar ss \bar d d uuu$ intermediate state is interesting because it can further proceed into
$\Lambda K^+\pi^+$ or to $pK^+\bar K^0$ or, via a narrow resonance $\phi$, to $p\phi\pi^+\to pK^+K^-\pi^+$.
These channels may balance the branching ratio and background conditions. In fact, a similar channel $pK^+K^-\pi^-$
has been used by E791 in the search for the neutral anti-charm pentaquark $P_{\bar cs}$~\cite{E791} but with the
trigger that four charged particles have the total zero charge. Here it must be +2. The ${\cal B}_c^+$ can decay into
three-prong final states $p\phi \to pK^+K^-$ or $\Lambda K^+$.

Returning to the production rate \ur{rate} it should be multiplied by a typical branching ratio $10^{-2}$ to a
particular observation channel, yielding a tiny observation rate of $10^{-9}$. Given that the total number of
events at LHC is $10^{15}/{\rm year}$, it still promises $10^6$ registrations of ${\cal B}_c$ per year. Respectively,
there could be as much as $10^5$ ${\cal B}_b$ events per year. At Fermilab the rate is 3 orders of magnitude
less but still probably accessible. It is interesting that a good fraction of ${\cal B}_b$ decays must be into
${\cal B}_c\;{\rm plus\;pions}$ since the dominant weak decay is $b\to cd\bar u$.

Because it is the main $b$-quark decay, ${\cal B}_c$ can be looked for at $B$-factories, Belle and Babar.
As a conservative estimate of the ${\cal B}_c$ production probability I would take the product of the probability
to create a charmed baryon of comparable mass (say, $\Sigma_c(2455)$ or $\Xi_c(2468)$), and of the probability
to create a deuteron. Searching for ${\cal B}_c$ in relativistic heavy ion collisions may be also promising since
the coalescence factor may be more favorable there.

\section{Conclusions}

If the number of colors $N_c$ is treated as a free algebraic parameter, baryon resonances are classified in
a simple way. At large $N_c$ all baryon resonances are basically determined by the ``intrinsic'' quark spectrum
which takes certain limiting shape at $N_c\to\infty$. This spectrum is the same in light baryons $q\ldots qq$
with $N_c$ light quarks $q$, and in heavy baryons $q\ldots qQ$ with $N_c\!-\!1$ light quarks and one heavy quark $Q$,
since the difference is a $1/N_c$ effect.

One can excite quark levels in various ways called either one-particle or particle-hole excitations; in both
cases the excitation energy is ${\cal O}(1)$. On top of each one-quark or quark-antiquark excitation there
is generically a band of $SU(3)$ multiplets of baryon resonances, that are rotational states of a baryon as
a whole. Therefore, the splitting between multiplets is ${\cal O}(1/N_c)$. The rotational band is terminated
when the rotational energy reaches ${\cal O}(1)$. Some multiplets which differ only by spin are degenerate
in the leading order but become split in the next ${\cal O}(1/N_c^2)$ order.

In reality $N_c$ is only 3, and the above idealistic hierarchy of scales is somewhat blurred. Nevertheless,
a close inspection of the spectrum of baryon resonances reveals certain hierarchy schematically summarized as follows:
\begin{itemize}
\item Baryon mass: ${\cal O}(N_c)$, numerically 1200 MeV, the average mass of the ground-state octet
\item One-quark and particle-hole excitations in the intrinsic spectrum: ${\cal O}(1)$, typically 400 MeV,
for example the excitation of the Roper resonance
\item Splitting between the centers of $SU(3)$ multiplets arising as rotational excitations of a given intrinsic state:
${\cal O}(1/N_c)$, typically 133 MeV
\item Splitting between the centers of rotational multiplets differing by spin, that are degenerate in the leading order:
${\cal O}(1/N_c^2)$, typically 44 MeV
\item Splitting inside a given multiplet owing to the nonzero strange quark mass: ${\cal O}(m_sN_c)$, typically 140 MeV.
\end{itemize}

In practical terms, the lowest light baryon multiplets $({\bf 8},{1/2}^+)$ and $({\bf 10},{3/2}^+)$ form the
``rotational band'' about the ground state, with the splitting between their centers being
$\frac{3}{2I_1}=230\,{\rm MeV}={\cal O}(1/N_c)$. The ground state of a heavy baryon (where one light quark is
replaced by a heavy one so that there is a hole in the light quarks valence shell) generates a rotational band of
three multiplets, $({\bf\bar 3},{1/2}^+)$, $({\bf 6},{1/2}^+)$ and $({\bf 6},{3/2}^+)$. These are precisely the
observed multiplets, and the prediction is that the two sextets are degenerate in the leading order whereas the
splitting between the ${\bf\bar 3}$ and ${\bf 6}$ is $\frac{1}{I_1}=153\,{\rm MeV}$. In reality the two sextets
are not degenerate but their splitting 67 MeV (an $1/N_c^2$ effect) is substantially less than the splitting
between the mean mass of the sextets and the anti-triplet, which is 162 MeV, off by only 6\% from the large-$N_c$
prediction.

This coincidence encourages to look what is the {\em lowest non-rotational excitation} of a heavy baryon
in the large-$N_c$ limit. Apparently, it is the particle-hole excitation where one takes an $s$ quark from
the highest filled shell and puts an $u$ or $d$ quark at the lowest $u,d$ valence shell, filling in the hole
there, see Fig.~4. The corresponding baryon resonances have the (penta) quark content ${\cal B}_c^{++}=cuud\bar s$,
${\cal B}_c^+=cudd\bar s$ with mass $m(\Lambda_c)+130\,{\rm MeV}=2420\,{\rm MeV}$ and ${\cal B}_b^+=buud\bar s$,
${\cal B}_b^0=budd\bar s$ with mass $m(\Lambda_b)+130\,{\rm MeV}=5750\,{\rm MeV}$. I call them ``Beta baryons''
(implying, of course, that ``Alpha baryons'' are the standard, mainly three-quark baryons). Actually,
${\cal B}_{b,c}$ baryons are part of the larger ${\bf\overline{15}}$ multiplet of pentaquarks, and there must
be three of them: two with spin-parity ${1/2}^+$ and one with ${3/2}^+$. The splitting of these
${\bf\overline{15}}$-plets is expected to be less than 100 MeV.

The arithmetic for the masses would be exact in the limit of infinite $N_c$, however in reality
${\cal O}(1/N_c)\sim 150\, {\rm MeV}$ corrections are allowed. However, there is still quite some room below
the threshold for strong decays, which is at 2780 MeV. Therefore, I believe that at least one but maybe two
or even three exotic pentaquarks ${\cal B}_{b,c}$ are stable with respect to strong decays. This makes their
discovery feasible, despite that the production rate is probably very low, see Section IX.

I think that the presented case for the heavy ${\cal B}_{b,c}$ pentaquarks is even stronger that it has been
for the \Th pentaquark~\cite{DPP-97}, whose mass I confirm here from a new, unified point of view. \\

I am grateful to Victor Petrov, Maxim Polyakov and Alexei Vladimirov for their help.
I thank Ben Mottelson and Semen Eidelman for useful discussions and Harry Lipkin for a correspondence.
This work has been supported in part by Russian Government grants RFBR-06-02-16786 and RSGSS-3628.2008.2.
I gratefully acknowledge discussions with many participants of the International workshop
{\it New Frontiers in QCD 2010}, Jan. 18 - Mar. 19, Kyoto, Japan.

\setcounter{section}{0}

\appendix
\section{Arbitrary $N_c$ prototypes of $SU(3)$ multiplets}

The usual $SU(3)$ multiplets, ${\bf 8}$ and ${\bf 10}$ for light baryons, ${\bf\bar 3}$ and ${\bf 6}$
for charmed and bottom baryons, arise for baryons made minimally of three quarks. If $N_c$ is considered
a free parameter baryons made of $N_c$ quarks fall into other $SU(3)$ multiplets. The aim of this Appendix
is to construct the generalizations of the ordinary multiplets to arbitrary $N_c$, such that the `prototype'
multiplets reduce to the usual ones at $N_c\!=\!3$.

A generic $SU(3)$ multiplet or irreducible representation
is uniquely determined by two non-negative integers $(p,q)$ having the meaning
of upper (lower) components of the irreducible $SU(3)$ tensor
$T^{\{f_1...f_p\}}_{\{g_1...g_q\}}$ symmetrized both in upper and lower
indices and with a contraction with any $\delta^{g_n}_{f_m}$ being zero.
Schematically, $q$ is the number of boxes in the lower line of the
Young tableau depicting an $SU(3)$  representation and $p$ is the
number of extra boxes in its upper line, see Fig.~7.

\begin{figure}[htb]
\begin{minipage}[]{.7\textwidth}
\includegraphics[width=\textwidth]{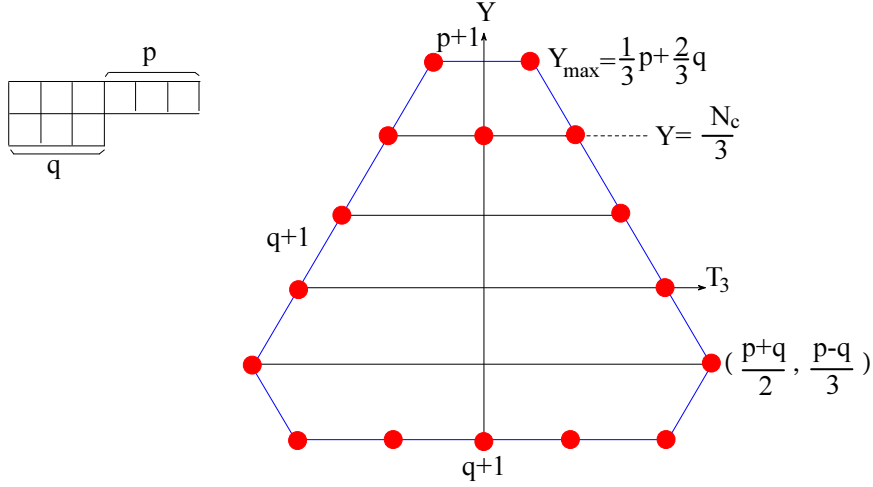}
\end{minipage}
\caption{A generic $SU(3)$ multiplet is, on the one hand, defined by the
Young tableau and on the other hand can be characterized by quantum numbers
$(T_3,Y)$ of its members filling a hexagon in the $(T_3,Y)$ axes (the weight
diagram).}
\label{fig:7}
\end{figure}

The dimension of a representation or the number of particles in the multiplet is
\beq
{\rm Dim}(p,q)=(p+1)(q+1)\left(1+\frac{p+q}{2}\right).
\la{Dim}\eeq
In the weight $(T_3,Y)$ diagram where $T_3$ is the third projection
of the isospin and $Y$ is the hypercharge, a generic
$SU(3)$ representation takes the form of a hexagon, whose the upper
horizontal side contains $p+1$ `dots' or particles, the adjacent
sides contain $q+1$ particles, with alternating $p+1$ and $q+1$
particles in the rest sides, the corners included --  see Fig.~7.
If either $p$ or $q$ is zero, the hexagon reduces to a triangle.

Particles at the upper (horizontal) side of the hexagon have the hypercharge
\beq
Y_{\rm max}=\frac{1}{3}\,p+\frac{2}{3}\,q
\la{Ymax}\eeq
being the maximal possible hypercharge of a multiplet with given $(p,q)$.
The number of particles in a horizontal line with given $Y$ is
\beq
n(Y)=\frac{4}{3}\,p+\frac{2}{3}\,q+1-Y.
\la{nY}\eeq
(possible non-unity multiplicities of particles with fixed $Y,\,T_3$ are neglected here.)

To find what $SU(3)$ multiplets are generated as rotation states from a given intrinsic
quark ground state or excitation, one has first to determine the hypercharge $Y'$
by counting the number of $u,d,s$ quarks involved in the intrinsic quark state,
\beq
Y'=\frac{1}{3}\,(\#{\rm \;of\;}u,d\;{\rm quarks})-\frac{2}{3}\,(\#{\rm \;of\;}s\;{\rm quarks}).
\la{Yprime}\eeq
For example, for the ground state light baryon $Y'=\frac{1}{3}\,N_c$ as there are $N_c$
$u,d$ quarks at the ground-state valence level, Fig.~1. The number of particles with this
hypercharge is related to the isospin $T'$ by the equation
\beq
2T'+1=n(Y')=\frac{4}{3}\,p+\frac{2}{3}\,q+1-Y'.
\la{Tp}\eeq

The logic of constructing the prototype multiplets is as follows. One first finds the allowed
multiplets that contain a given $Y'$ for $N_c\!=\!3$, and reads off the $T'$ for those multiplets
from the weight diagrams. By construction, $Y'\leq Y_{\rm max}$, the highest possible hypercharge of
an allowed multiplet. Let us write $Y_{\rm max}=Y'+X$ where $X$ is the number of steps in the
weight diagram, by which the top of the diagram is separated from $Y'$.

In principle, there are many ways how to generalize the real-world $SU(3)$ multiplets to arbitrary
$N_c$. The natural one~\cite{DuPrasz,DP-04} is to fix at all $N_c$ the shape of the weight diagram
at its upper part, meaning fixing $T'$ and $X$ for all $N_c$ as they appear at $N_c\!=\!3$. Physically,
it corresponds to the generalization where one adds more $u,d$ quarks to the baryon as one increases
$N_c$, and not $s$ quarks. The $(p,q)$ numbers of the prototype multiplet in question is then found
from \Eq{Tp} and from
\beq
Y'+X=Y_{\rm max}=\frac{1}{3}\,p+\frac{2}{3}\,q.
\la{2ndpq}\eeq
The rotational energy of the prototype multiplet is given by \Eq{Erot}. Let us consider
several examples of building the prototype multiplets at arbitrary $N_c$. We assume $N_c$ is odd
such that baryons are fermions.\\

\noindent
\underline{Light baryons, ground state}\\

In this case $Y'=1$ at $N_c\!=\!3$, and the allowed multiplets are ${\bf 8},\;{\bf 10}$ since these
multiplets contain a line with $Y'=1$; for the octet there are two such particles, hence $T'=\frac{1}{2}$,
whereas for the decuplet there are four such particles, hence $T'=\frac{3}{2}$. In both cases it is the upper
line, therefore $X=0$.

Generalizing these multiplets to arbitrary $N_c$ we fix $T'$ and $X$ what they are at $N_c\!=\!3$ but change
$Y'$ according to \Eq{Yprime}. For the ground state $Y'=\frac{N_c}{3}$. Solving \Eqs{Tp}{2ndpq} with respect
to $(p,q)$ we find the prototype `octet':
\beq\n
\left\{\begin{array}{c}p\\q\\{\rm Dim}\\E_{\rm rot}\end{array}\right.\quad =\quad
\left\{\begin{array}{c}1\\ \frac{N_c-1}{2}\\ \frac{(N_c+1)(N_c+5)}{4}\\ \frac{N_c}{4I_2}+\frac{3}{8I_1}\end{array}\right.
\quad \stackrel{N_c\to 3}{\longrightarrow}\quad
\left\{\begin{array}{c}1\\1 \\ {\bf 8} \\ \frac{3}{4I_2}+\frac{3}{8I_1}.\end{array}\right.
\eeq
The prototype `decuplet' is
\beq\n
\left\{\begin{array}{c}p\\q\\{\rm Dim}\\E_{\rm rot}\end{array}\right.\quad =\quad
\left\{\begin{array}{c}3\\ \frac{N_c-3}{2}\\ \frac{(N_c-1)(N_c+7)}{4}\\ \frac{N_c}{4I_2}+\frac{15}{8I_1}\end{array}\right.
\quad \stackrel{N_c\to 3}{\longrightarrow}\quad
\left\{\begin{array}{c}3\\0 \\ {\bf 10} \\ \frac{3}{4I_2}+\frac{15}{8I_1}.\end{array}\right.
\eeq
The rotational energies of the prototype ``${\bf 8}$'' and ``${\bf 10}$'' differ by $\frac{3}{2I_1}={\cal O}(1/N_c)$.
The term $\frac{N_c}{4I_2}={\cal O}(1)$ is a common shift in this (and subsequent) examples.
Higher multiplets containing $Y'=\frac{N_c}{3}$ have ${\cal O}(1)$ rotational splitting and should
be discarded for this reason.

The spins of these two prototype multiplets are found from the vector addition rule \ur{Tprime}. In this case $J_1=J_2=0$,
hence the spin $J=\frac{1}{2}$ for the `octet', and $J=\frac{3}{2}$ for the `decuplet'. \\

\noindent
\underline{Light baryons, $0^+\to{\frac{1}{2}}^-$ transition}\\

This is the intrinsic one-quark excitation {\it 1} in Fig.~2. At $N_c\!=\!3$ it corresponds to $Y'=0$, and the allowed
multiplets are, in principle,  the singlet, the octet and the decuplet, with $T'=0,\,(0,1),\,1$, respectively,
and $X=1$ in all cases. The generalization changes $Y'=\frac{N_c-3}{3}$. The prototype `singlet' is
\beq\n
\left\{\begin{array}{c}p\\q\\{\rm Dim}\\E_{\rm rot}\end{array}\right.\quad =\quad
\left\{\begin{array}{c}0\\ \frac{N_c-3}{2}\\ \frac{(N_c-1)(N_c+1)}{8}\\ \frac{N_c-3}{4I_2}\end{array}\right.
\quad \stackrel{N_c\to 3}{\longrightarrow}\quad
\left\{\begin{array}{c}0\\0 \\ {\bf 1} \\ 0.\end{array}\right.
\eeq
The spin of this `singlet' is $J=\frac{1}{2}$.

To build the `octet' we take $T'=1$ and $X=1$ and find
\beq\n
\left\{\begin{array}{c}p\\q\\{\rm Dim}\\E_{\rm rot}\end{array}\right.\quad =\quad
\left\{\begin{array}{c}1\\ \frac{N_c-1}{2}\\ \frac{(N_c+1)(N_c+5)}{4}\\ \frac{N_c-2}{2I_2}+\frac{1}{I_1}\end{array}\right.
\quad \stackrel{N_c\to 3}{\longrightarrow}\quad
\left\{\begin{array}{c}1\\1 \\ {\bf 8} \\ \frac{1}{2I_2}+\frac{1}{I_1}.\end{array}\right.
\eeq
We note that the rotational energy differs from that of the `singlet' by ${\cal O}(1)$. Therefore, this multiplet,
strictly speaking, is {\em not} a rotational excitation of the intrinsic state. In this case the rotational band
consists of only one state, the `singlet'. At $N_c\!=\!3$ it is the $\Lambda(1405,1/2^-)$. \\

\noindent
\underline{Light baryons, ${\frac{1}{2}}^+\to{\frac{1}{2}}^-$ transition}\\

This is the intrinsic particle-hole excitation {\it 2} in Fig.~2. $Y'$ is the same as for the ground state
(see above), hence the rotational band consists of `octets' and `decuplets'. Their spins are found from \Eq{Tprime}:
the rotational band about this transition consists of the following multiplets: $({\bf 8},1/2^-)$ (twice),
$({\bf 8},3/2^-)$, $({\bf 10},1/2^-)$, $({\bf 10},3/2^-)$ (twice) and $({\bf 10},5/2^-)$. In the leading $1/N_c$
order the three octets are degenerate and so are the four decuplets.

In reality, there is indeed a well-grouped triad of octets with the approximate centers at $1615({\bf 8},1/2^-)$,
$1710({\bf 8},1/2^-)$ and $1680({\bf 8},3/2^-)$. Given that degeneracy is always lifted by any additional interaction
(for example by the strange quark mass) such that one state goes up and the other goes down, this seems to be
a success of the description. The situation is worse with parity-minus decuplets. Two of them are rather well
identified at approximately $1758({\bf 10},1/2^-)$ and $1850({\bf 10},3/2^-)$. As to the rest decuplets, there
is only a one-star $\Delta(1940,3/2^-)$ and a three-star $\Delta(1930,5/2^-)$ in the PDG, which can fit into
the picture but the experimental situation is inconclusive. \\

\noindent
\underline{Light baryons, ${\frac{1}{2}}^+\to 0^+$ transition}\\

This is the intrinsic particle-hole excitation {\it 4} in Fig.~2. At $N_c\!=\!3$ it corresponds to $Y'=2$, and the allowed
multiplets are the ${\bf\overline{10}}$ and the ${\bf 27}$, with $T'=0,\,1$, respectively, and $X=0$ in both cases.
The generalization is $Y'=\frac{N_c+3}{3}$. The prototype `anti-decuplet' is
\beq\n
\left\{\begin{array}{c}p\\q\\{\rm Dim}\\E_{\rm rot}\end{array}\right.\quad =\quad
\left\{\begin{array}{c}0\\ \frac{N_c+3}{2}\\ \frac{(N_c+5)(N_c+7)}{8}\\ \frac{N_c+3}{4I_2}\end{array}\right.
\quad \stackrel{N_c\to 3}{\longrightarrow}\quad
\left\{\begin{array}{c}0\\3 \\ {\bf\overline{10}} \\ \frac{3}{2I_2}.\end{array}\right.
\eeq
The spin of this `${\bf\overline{10}}$' is $J=\frac{1}{2}$.

To build the prototype `${\bf 27}$'-plet we put $T'=1$ and $X=0$ and find
\beq\n
\left\{\begin{array}{c}p\\q\\{\rm Dim}\\E_{\rm rot}\end{array}\right.\quad =\quad
\left\{\begin{array}{c}2\\ \frac{N_c+1}{2}\\ \frac{3(N_c+3)(N_c+9)}{8}\\ \frac{N_c+3}{4I_2}+\frac{1}{I_1}\end{array}\right.
\quad \stackrel{N_c\to 3}{\longrightarrow}\quad
\left\{\begin{array}{c}2\\2 \\ {\bf 27} \\ \frac{3}{2I_2}+\frac{1}{I_1}.\end{array}\right.
\eeq
The spins of the `${\bf 27}$'-plet are found from the vector addition law ${\bf J}={\bf 1}+{\bf\frac{1}{2}}$, and
can be both $J=\frac{1}{2}$ and $J=\frac{3}{2}$. In the leading order in $1/N_c$ they are degenerate but both separated
by $\frac{1}{I_1}\approx 150\,{\rm MeV}$ from the `${\bf\overline{10}}$'. In practical terms it means that there
must be an exotic triplet $\Theta^{++},\Theta^+,\Theta^0$ with spins $\frac{1}{2}$ and $\frac{3}{2}$ about 150 MeV
heavier that the singlet $\Theta^+$ belonging to the anti-decuplet. \\

\noindent
\underline{Heavy baryons, ground state}\\

In this case $Y'=\frac{2}{3}$ at $N_c\!=\!3$, and the allowed multiplets are ${\bf\bar 3},\;{\bf 6}$ since these
multiplets contain particles with $Y'=\frac{2}{3}$; for the ${\bf\bar 3}$ there is one such particle, hence $T'=0$,
whereas for the ${\bf 6}$ there are three such particles, hence $T'=1$. In both cases it is the upper
line, therefore $X=0$.

Generalizing these multiplets to arbitrary $N_c$ we fix $T'$ and $X$ what they are at $N_c\!=\!3$ but change
$Y'$ according to \Eq{Yprime}. For the ground state $Y'=\frac{N_c-1}{3}$. Solving \Eqs{Tp}{2ndpq} with respect
to $(p,q)$ we find the prototype `anti-triplet':
\beq\n
\left\{\begin{array}{c}p\\q\\{\rm Dim}\\E_{\rm rot}\end{array}\right.\quad =\quad
\left\{\begin{array}{c}0\\ \frac{N_c-1}{2}\\ \frac{3(N_c+1)(N_c+3)}{8}\\ \frac{N_c-1}{4I_2}\end{array}\right.
\quad \stackrel{N_c\to 3}{\longrightarrow}\quad
\left\{\begin{array}{c}0\\1 \\ {\bf\bar 3} \\ \frac{1}{2I_2}.\end{array}\right.
\eeq
Its spin is determined from ${\bf J}={\bf T'}+{\bf J_h}$ and can be only $\frac{1}{2}$.

The prototype `sextet' is
\beq\n
\left\{\begin{array}{c}p\\q\\{\rm Dim}\\E_{\rm rot}\end{array}\right.\quad =\quad
\left\{\begin{array}{c}2\\ \frac{N_c-3}{2}\\ \frac{3(N_c-1)(N_c+5)}{8}\\ \frac{N_c-1}{4I_2}+\frac{1}{I_1}\end{array}\right.
\quad \stackrel{N_c\to 3}{\longrightarrow}\quad
\left\{\begin{array}{c}2\\0 \\ {\bf 6} \\ \frac{1}{2I_2}+\frac{1}{I_1}.\end{array}\right.
\eeq
The spins are $\frac{1}{2}$ and $\frac{3}{2}$. \\

\noindent
\underline{Heavy baryons, $0^+\to{\frac{1}{2}}^-$ transition}\\

This is the intrinsic one-quark excitation analogous to transition {\it 1} in Fig.~2.
At $N_c\!=\!3$ it corresponds to $Y'=-\frac{1}{3}$, and the allowed
multiplets are the anti-triplet and sextet with $T'=\frac{1}{2}$ and $X=1$ in both cases.
The generalization is $Y'=\frac{N_c-4}{3}$. The prototype `anti-triplet' is
\beq\n
\left\{\begin{array}{c}p\\q\\{\rm Dim}\\E_{\rm rot}\end{array}\right.\quad =\quad
\left\{\begin{array}{c}0\\ \frac{N_c-1}{2}\\ \frac{3(N_c+1)(N_c+3)}{8}\\ \frac{2N_c-5}{4I_2}+\frac{3}{8I_1}\end{array}\right.
\quad \stackrel{N_c\to 3}{\longrightarrow}\quad
\left\{\begin{array}{c}0\\1 \\ {\bf\bar 3} \\ \frac{1}{4I_2}+\frac{3}{8I_1}.\end{array}\right.
\eeq
We see that already the lowest possible multiplet has an unacceptable ${\cal O}(1)$ energy as compared to the
ground state rotational energy. It means that from the large-$N_c$ viewpoint the existence of this multiplet
strictly speaking cannot be claimed. One has to explain the parity-minus heavy baryons by other means -- see Section V.\\

\noindent
\underline{Heavy baryons, ${\frac{1}{2}}^+\to 0^+$ transition}\\

This is the intrinsic particle-hole excitation depicted in Fig.~4, left.
At $N_c\!=\!3$ it corresponds to $Y'=\frac{5}{3}$, and the allowed
multiplet is the anti-decapenta-plet shown in Fig.~4, right, with $T'=\frac{1}{2}$ and $X=0$.
Its arbitrary-$N_c$ generalization has $Y'=\frac{N_c+2}{3}$. The prototype `anti-decapenta-plet' is characterized by
\beq\n
\left\{\begin{array}{c}p\\q\\{\rm Dim}\\E_{\rm rot}\end{array}\right.\quad =\quad
\left\{\begin{array}{c}1\\ \frac{N_c+1}{2}\\ \frac{(N_c+3)(N_c+7)}{4}\\ \frac{N_c+2}{4I_2}+\frac{3}{8I_1}\end{array}\right.
\quad \stackrel{N_c\to 3}{\longrightarrow}\quad
\left\{\begin{array}{c}1\\2 \\ {\bf\overline{15}} \\ \frac{5}{4I_2}+\frac{3}{8I_1}.\end{array}\right.
\eeq
The spins of the `anti-decapenta-plet' are found from the relation ${\bf J}={\bf \frac{1}{2}}+{\bf \frac{1}{2}}
+{\bf \frac{1}{2}}$. Therefore, there are two multiplets with spin $\frac{1}{2}$ and one multiplet with spin $\frac{3}{2}$,
all degenerate in the leading order in $1/N_c$. Their lightest members are the exotic Beta-baryons ${\cal B}_{b,c}$,
the main prediction of this paper.


\begin{thebibliography}{99}
\bibitem{footnote-4}
On the naming: After I proposed the name \Th for the pentaquark $uudd\bar s$, its heavy counterparts $uudd\bar c$
and $uudd\bar b$ conjectured by Karliner and Lipkin~\cite{KarLip} have been named $\Theta_c$ and $\Theta_b$, respectively,
according to the tradition to denote baryons with subscripts $c,b$ when the strange quark in a light baryon is replaced
by a heavy one. Earlier pentaquarks of the type $qqqs\bar Q$ have been hypothesized by Gignoux, Silvestre-Brac and
Richard~\cite{Gignoux} and Lipkin~\cite{Lipkin} and denoted as $P_{\bar cs}$. I propose here a very different type
of pentaquarks $qqqQ\bar s$ which I suggest to call ``Beta baryons'' and denote as ${\cal B}_{c,b}$ (calligraphic `Bee'
in LaTeX). The implication is that ``Alpha baryons'' are mainly the three-quark ones, of course.
\end{thebibliography}
\end{document}